\documentclass[12pt]{article}
\usepackage[utf8]{inputenc}
\usepackage{graphicx} 
\usepackage{dcolumn} 
\usepackage{bbm}
\usepackage{bm}
\usepackage{dsfont}
\usepackage{amsfonts}
\usepackage{amsthm}
\usepackage{amsmath}
\usepackage{amssymb}
\usepackage{epsfig}
\usepackage{color}
\usepackage{textcomp}
\usepackage{hyperref}
\usepackage{titlesec}
\usepackage{slashed}
\usepackage{caption}
\usepackage{subcaption}
\usepackage{booktabs}
\usepackage{multirow}
\usepackage{cite}
\usepackage{placeins}
\usepackage{comment}
\usepackage{braket}
\usepackage{authblk}
\usepackage{physics}
\usepackage{lmodern}
\numberwithin{equation}{section}

\title{\textbf{Graviton-Photon Oscillations as a Probe of Quantum Gravity}}
\author{Andrea Palessandro \thanks{apalessandro@deloitte.com}}
\affil{\small Deloitte AI Institute}
\date{}

\begin{document}

\maketitle

\begin{abstract}
    \noindent The Gertsenshtein effect could in principle be used to detect a single graviton by firing it through a region filled with a constant magnetic field that enables its conversion to a photon, which can be efficiently detected via standard techniques. The quantization of the gravitational field could then be inferred indirectly. We show that for currently available single-photon detector technology, the Gertsenshtein detector is generically inefficient, meaning that the probability of detection is $\ll 1$. The Gertsenshtein detector can become efficient on astrophysical scales for futuristic single-photon detectors sensitive to frequencies in the Hz to kHz range. It is not clear whether such devices are in principle possible. 
\end{abstract}

\section{Introduction}
To this day, more than a century after the birth of General Relativity, physics has yet to produce a satisfactory theory of quantum gravity. Given the difficulties in quantizing gravity, one might ask whether gravity is in fact quantized and, if it is, whether there is any experimental way to detect its quantization. 

In a famous paper \cite{Bohr}, Bohr and Rosenfeld showed that the quantum-mechanical limitation of measurement on the motion of test masses implies an analogous limitation in the measurement of electromagnetic field-strengths, and so a classical electromagnetic field interacting with a quantum-mechanical measuring apparatus inevitably leads to inconsistencies. Freeman Dyson \cite{Dyson}, following Bronstein \cite{Bronstein}, pointed out that, due to its universally attractive nature, the same is not true of gravity: it is perfectly consistent to have a classical gravitational field interacting with quantum matter.

There is a long tradition of (thought) experiments attempting to prove the quantum nature of gravity in the spirit of Bohr and Rosenfeld \cite{Feynman, Hannah, Page, DePalma}, but, on closer inspection, they all ultimately fail to compel such a quantization \cite{Mattingly, Baym, Rydving}, with the possible exception of gravitationally mediated entanglement \cite{Belenchia, Danielson}. The basic reason is that gravity introduces a minimum length\footnote{Certain theories of emergent or asymptotically safe gravity do not exhibit a minimal length \cite{Hossenfelder}.} which acts as a fundamental limit to any distance measurement \cite{Calmet}. Trying to resolve finer distances results in black hole production, with any excess energy further increasing the minimum region of momentum uncertainty. The experiments that have been proposed so far, both imagined and real, all end up relying on trans-planckian resolution, and this is why they fail to demonstrate quantization. The cosmological argument of Krauss and Wilczek \cite{Krauss} also relies on trans-planckian physics, as I will explain in \S \ref{relation}, and is therefore plagued by similar problems.

If the quantization of the gravitational field is not a logical necessity, it can only be established empirically. Conceptually, the most immediate way to do that is to detect a single graviton. Dyson was the first to seriously investigate the feasibility of graviton detection and found it to be impossible in practice due to gravitational collapse \cite{Dyson}. Rothman and Boughn later provided detailed arguments and calculations in support of Dyson's conclusion \cite{Rothman, Boughn}.

One of the many graviton detector architectures proposed by Dyson utilizes the conversion of gravitons to photons, known as the Gertsenshtein effect after its discoverer \cite{Gertsenshtein}. In this experiment, a single graviton is incident on a region filled with a constant magnetic field, which couples the gravitational and electromagnetic degrees of freedom \cite{Ejlli, Carney}. In this region, the graviton can convert to a photon, which can then be detected with high probability using traditional photon counting technology. The question we ask in this work is the following:
\begin{quote}
    Using a detector architecture based on the Gertsenshtein effect, is it possible, in principle, to \textit{reliably} detect a single graviton?
\end{quote}
We tentatively conclude that the answer to this question is no for currently existing single-photon detectors, if \textit{reliably} is taken to mean ``with probability one". Futuristic single-photon detector technology sensitive to frequencies many orders of magnitude below the current near infrared limit (roughly in the range from Hz to kHz) could in principle allow the efficient detection of single gravitons using a Gertsenshtein architecture. However, it is not clear whether (single) photons with such long wavelengths can ever be efficiently detected either.

Interestingly, the inefficiency of the Gertsenshtein detector is ultimately due to nonlinear electromagnetic vacuum polarization effects, which destroy the coherence of graviton-photon mixing below cosmological scales. A consistent treatment of graviton-photon oscillations in an expanding universe shows that these always lie beyond the Hubble radius and are in principle unobservable \cite{Anninos}.

The paper is structured in the following way: in \S \ref{non-linear} we write down the general coupled gravito-electromagnetic equations in a polarized vacuum using the formalism of \cite{Pal}. In \S \ref{mixing} we derive the mixing matrix between graviton and photon quantum states and calculate their transition probability. In \S \ref{weak} we study the Gertsenshtein effect in the weak field limit  and make contact with previous results. In \S \ref{strong} we study the effect in the strong field regime and explain why coherent oscillations must stop above the Schwinger limit \cite{Zeldovich}. In \S \ref{experiment} we discuss in details a simple experiment based on the Gertsenshtein effect that could detect a single graviton, and conclude that the experiment cannot realistically be performed in our universe, in the sense specified above. In \S \ref{relation} we compare the Gertsenshtein detector with other graviton detector architectures present in the literature, and find that the reason why the Gertsenshtein detector is inefficient is qualitatively different than in all other cases.

\section{Graviton-photon oscillations in the non-linear regime}\label{non-linear}
In this section we study the classical Gertsenshtein effect in the non-linear electromagnetic regime using the formalism of \cite{Pal}. Specifically, we include self-interactions of the electromagnetic field to one-loop order. We work with natural units $\varepsilon_0 = \hbar = c = 1$, and metric signature $(-,+,+,+)$. Here and below $T_{\mu \nu}$ is the electromagnetic stress-energy tensor, $F_{\mu \nu} = \partial_\mu A_\nu - \partial_\nu A_\mu$ the electromagnetic field tensor, $\Tilde{F}_{\mu \nu} = 1/2 \, \epsilon^{\mu \nu \alpha \beta} F_{\alpha \beta}$ its dual, $\alpha \equiv q^2/4\pi$ the fine structure constant, $q$ the electron charge, $m_e$ the electron mass, and $G \equiv m_p^{-2}$ Newton's constant written in terms of the Planck mass $m_p$.

For small metric perturbations 
\begin{equation}\label{metric}
    g_{\mu \nu} = \eta_{\mu \nu} + h_{\mu \nu}, \quad |h_{\mu \nu}| \ll 1,
\end{equation}
the Lagrangian density describing the interaction between the gravitational and electromagnetic fields is 
\begin{equation}\label{lagrangian}
    \mathcal{L} = \mathcal{L}_G + \mathcal{L}_{\text{EM}} + \mathcal{L}_I ,
\end{equation}
where 
\begin{equation}
    \mathcal{L}_G = -\frac{1}{16 \pi G}\left( \frac{1}{2} \partial_\mu h_{\alpha \beta}\partial^\mu h^{\alpha \beta} - \frac{1}{2} \partial_\mu h \partial^\mu h + \partial_\mu h^{\mu \nu} \partial_\nu h - \partial_\mu h^{\mu \nu} \partial_\rho h^\rho_\nu\right)
\end{equation}
is the kinetic term of the gravitational field,
\begin{equation}\label{LEM}
    \mathcal{L}_{\text{EM}} = -\frac{1}{4} F_{\mu \nu}F^{\mu \nu} + \mathcal{L}^{(1)}_{EM}
\end{equation}
is the kinetic term of the electromagnetic field, which includes vacuum polarization effects to one loop order, and
\begin{equation}\label{LI}
    \mathcal{L}_I = \frac{1}{2} h_{\mu \nu} T^{\mu \nu}
\end{equation}
is the linearized interaction term.

The electric and magnetic fields in vacuum\footnote{More precisely, when vacuum polarization effects are negligible.} are defined in terms of the field tensor as $E_i \equiv F_{i0}$ and $B_i \equiv \epsilon_{ijk}F^{jk}/2$. With the aid of the gauge and Lorentz invariants
\begin{align}\label{SP}
\begin{split}
   S &= -\frac{1}{4} F_{\mu \nu}F^{\mu \nu} = \frac{1}{2} (E^2 - B^2), \\
    P &= -\frac{1}{4}F_{\mu \nu}\Tilde{F}^{\mu \nu} = \vec{E}\cdot \vec{B},
\end{split}
\end{align}
we can write the one-loop correction to the classical QED Lagrangian as \cite{Weisskopf}
\begin{equation}\label{oneloop}
    \mathcal{L}_{EM}^{(1)}(a,b) = - \frac{m_e^4}{8 \pi^2} \int_0^\infty ds \frac{e^{-s}}{s^3} \left[ (as)(bs) \cot(as) \coth(bs) -1 +\frac{s^2}{3} (a^2-b^2)\right],
\end{equation}
where
\begin{equation}\label{ab}
    a \equiv \frac{q}{m_e^2} \sqrt{\sqrt{S^2+P^2} + S} = \frac{E}{E_c}, \quad b \equiv \frac{q}{m_e^2}\sqrt{\sqrt{S^2+P^2} - S} = \frac{B}{B_c}.
\end{equation}
In natural units $E_c = B_c \equiv m_e^2/q$ is the Schwinger limit, above which the electromagnetic field becomes nonlinear. The Lagrangian (\ref{oneloop}) encodes the quantum corrections to photon propagation due to the creation of virtual electron-positron pairs from the vacuum when background electromagnetic fields are present. We assume that quantum gravity effects emerge at the Planck scale, and thus disregard any analogous contribution due to gravitational physics beyond tree level, given that it would be suppressed by powers of $m_p^2$. Were this not true, for example if large extra dimensions existed, one would need to take explicitly into account the nonlinearities of the gravitational field as well. For details, see appendix \ref{remarks}.

The coupled gravito-electromagnetic equations are then given by the least action principle:
\begin{align}\label{coupled}
    \begin{split}
        \frac{\partial \mathcal{L}}{\partial h_{\mu\nu}} - \partial_\alpha \frac{\partial \mathcal{L}}{\partial (\partial_\alpha h_{\mu\nu})} &= 0, \\
        \frac{\partial \mathcal{L}}{\partial A_\nu} - \partial_\mu \frac{\partial \mathcal{L}}{\partial (\partial_\mu A_\nu)} &= 0.
    \end{split}
\end{align}
Given the Lagrangian (\ref{lagrangian}), the equation of motion for the metric perturbation $h_{\mu \nu}$ is
\begin{equation}
    \frac{1}{2}\left( \partial_\sigma \partial_\mu h^\sigma_\nu + \partial_\sigma \partial_\nu h^\sigma_\mu - \partial_\mu \partial_\nu h - \Box h_{\mu \nu} - \eta_{\mu \nu} \partial_\rho \partial_\lambda h^{\rho \lambda} + \eta_{\mu \nu} \Box h\right) = 8 \pi G T_{\mu \nu}.
\end{equation}
In the transverse-traceless (TT) gauge, $\partial_\mu h^{\mu \nu} = h^\mu_\mu = 0$, and the equation simplifies to\footnote{The d'Alembertian form of the gravitational wave equation is valid for flat, static spacetimes only. In the more general case of a curved background, a covariant approach like the one used in \cite{Anninos2} or \cite{Tsagas} must be employed.}
\begin{equation}\label{boxh}
    \Box h_{\mu \nu} = - 16 \pi G T_{\mu \nu}.
\end{equation}
Using the chain rule, the equation of motion for the electromagnetic field can be written as
\begin{equation}\label{EL}
    - \frac{1}{2}\partial_\mu \left( \frac{\partial \mathcal{L}_{EM}}{\partial F_{\alpha \beta}} \frac{\partial F_{\alpha \beta}}{\partial \partial_\mu A_\nu}  \right) = - \partial_\mu \left(\frac{\partial \mathcal{L}_{EM}}{\partial F_{\mu \nu}}\right) = 0.
\end{equation}
Defining the dielectric tensor \cite{Lifshitz}
\begin{equation}\label{Gdef}
    \mathcal{G}^{\mu \nu} \equiv - \frac{\partial \mathcal{L}_{EM}}{\partial F_{\mu \nu}},
\end{equation}
we can write (\ref{EL}) in compact form as
\begin{equation}\label{Gtilde}
    \partial_\mu \mathcal{G}^{\mu \nu} = 0.
\end{equation}
The Bianchi identity still holds:
\begin{equation}\label{Ftilde}
    \partial_\mu \Tilde{F}^{\mu \nu} = 0.
\end{equation}
Equations (\ref{Gtilde}) and (\ref{Ftilde}) are the dynamical equations for the electromagnetic field (linearly) coupled to gravity when vacuum polarization effects cannot be neglected.

The physical fields in the presence of vacuum polarization are $\Vec{D}$ and $\Vec{H}$, and can be defined in terms of the dielectric tensor $\mathcal{G}^{\mu \nu}$ in complete analogy with the Maxwell case:
\begin{align}\label{DH}
    \begin{split}
        D_i &\equiv \mathcal{G}_{i0} = - \frac{\partial \mathcal{L}_{EM}}{\partial F^{i0}} = \frac{\partial \mathcal{L}_{EM}}{\partial E^i}, \\
        H_i &\equiv \frac{1}{2} \epsilon_{ijk} \mathcal{G}^{jk} = - \frac{1}{2} \epsilon_{ijk} \frac{\partial \mathcal{L}_{EM}}{\partial F^{jk}} = - \frac{\partial \mathcal{L}_{EM}}{\partial B^i}.
    \end{split}
\end{align}
Note that at low energies one can ignore the nonlinear terms, so $\mathcal{L}_{EM} = (E^2-B^2)/2$, and the equations above reduce to $\Vec{E} = \partial{\mathcal{L}_{EM}}/\partial \Vec{E}$ and $\Vec{B} = - \partial{\mathcal{L}_{EM}}/\partial \Vec{B}$, as they should.

We now assume that $h_{\mu \nu}$ describes a gravitational wave with two independent polarizations $h_{11} = - h_{22}$ and $h_{12} = h_{21}$. Furthermore, we assume that both the gravitational and electromagnetic waves are propagating in the $z$ direction, and that in addition to the internal electric and magnetic fields of the electromagnetic wave $\Vec{e} = e(z,t) \hat{x}$ and $\Vec{b} = b(z,t) \hat{y}$, there is a background magnetic field $\Vec{B_0} = B_0 \hat{x}$ such that $\Vec{B} = \Vec{b} + \Vec{B_0}$, with $B_0 \gg b,e$. Note here that we are ignoring gravitational backreaction on the grounds that higher order corrections to graviton propagation are suppressed by powers of the Planck mass, as explained in appendix \ref{remarks}.

In this setup, we can write (\ref{DH}) as $\Vec{D} = \Vec{E} + \Vec{\Pi}$ and $\Vec{H} = \Vec{B} + \Vec{M}$, where
\begin{align}\label{PM}
\begin{split}
    \Vec{\Pi} = \frac{q^2}{a m_e^4}\frac{\partial \mathcal{L}^{(1)}_{EM}}{\partial a} \Vec{E},\\
    \Vec{M} = -\frac{q^2}{b m_e^4}\frac{\partial \mathcal{L}^{(1)}_{EM}}{\partial b} \Vec{B},
\end{split}
\end{align}
are the vacuum polarization and magnetization vectors, respectively. Thus, given (\ref{DH}) and (\ref{PM}) we can very generally incorporate vacuum polarization effects in our setup by redefining the physical fields as $\Vec{D} \equiv \varepsilon \Vec{E}$ and $\Vec{H} \equiv \mu^{-1} \Vec{B}$, with
\begin{align}\label{emu}
    \begin{split}
        \varepsilon =  1 + \frac{q^2}{a m_e^4}\frac{\partial \mathcal{L}^{(1)}_{EM}}{\partial a}, \\
        \mu^{-1} = 1 - \frac{q^2}{b m_e^4}\frac{\partial \mathcal{L}^{(1)}_{EM}}{\partial b},
    \end{split}
\end{align}
the electric permittivity and magnetic permeability of the vacuum, respectively. These deviate from their classical values as a result of one loop QED contributions to vacuum polarization.

We have kept the metric implicit in (\ref{Gtilde}). We can make it explicit by writing it in covariant form as
\begin{equation}
   \partial_\mu \left( g^{\alpha \mu} g^{\beta \nu} \mathcal{G}_{\alpha \beta}\right) = 0.
\end{equation}
For small metric perturbations $g^{\mu \nu} = \eta^{\mu \nu} - h^{\mu \nu}$, therefore, and to first order in $h_{\mu \nu}$,
\begin{equation}\label{Gmnh}
    \partial_\mu \mathcal{G}^{\mu \nu} - \mathcal{G}^\mu_{\,\, \beta} \partial_\mu h^{\beta \nu}= 0.
\end{equation}
For $\nu = 0$, (\ref{Gmnh}) then gives $\Vec{\nabla} \cdot \Vec{D} = 0$. For $\nu = i$, it gives\footnote{From here on we use the comma notation for partial derivatives, e.g. $\partial_\mu f_\nu \equiv f_{\nu,\mu}$.}
\begin{equation}\label{Di}
    \dot{D}_i - \epsilon_{ijk} H_{k,j}  - \dot{h}_{ij}D_j + \epsilon_{jkl} H_l h_{ij,k} = 0.
\end{equation}
Note that for $h_{ij} = 0$ the equation above reduces to the flat space result $\dot{\Vec{D}} = \Vec{\nabla}\cross \Vec{H}$.
Taking the curl of (\ref{Di}) gives
\begin{equation}\label{curl}
    \epsilon_{ijk} \dot{D}_{k,j} - \epsilon_{ijk} \epsilon_{klm} H_{m,lj}  - \epsilon_{ijk} \dot{h}_{kl}D_{l,j} + \epsilon_{ijk}\epsilon_{lmn} (H_n h_{kl,m})_{,j} = 0.
\end{equation}
The Bianchi identity (\ref{Ftilde}) gives the homogeneous Maxwell's equations $\Vec{\nabla} \cdot \Vec{B} = 0$, and $\dot{\Vec{B}} = - \Vec{\nabla} \cross \Vec{E}$. Given $\Vec{D} \equiv \varepsilon \Vec{E}$ and $\Vec{H} \equiv \mu^{-1} \Vec{B}$, we can write these in terms of the physical fields as 
\begin{align}\label{homoHD}
    \begin{split}
        &\Vec{\nabla} \cdot \Vec{H} = 0, \\
        &\dot{\Vec{H}} = - v^2 (\Vec{\nabla} \cross \Vec{D}),
    \end{split}
\end{align}
where $v^2 \equiv (\varepsilon \mu)^{-1}$ is the group velocity of the electromagnetic radiation.

We can use (\ref{homoHD}) to write the equation of motion for the propagating field $b$ by eliminating any dependency on $e$ in (\ref{curl}). For $B_0 \gg b,e$, this gives
\begin{equation}\label{bdiff}
    \Ddot{b} - v^2 b'' = v^2 h''_{12} B_0.
\end{equation}

To write down the corresponding equation for the gravitational mode, we need to work out the electromagnetic stress-energy tensor. This is defined as
\begin{equation}\label{Tmn}
   4 \pi  T^{\mu \nu} = g^{\mu \nu} \mathcal{L}_{EM} - \frac{\partial \mathcal{L}_{EM}}{\partial (\partial_\mu A_\alpha)}\partial^\nu A_\alpha = \mathcal{G}^{\mu \alpha} F^\nu_{\, \, \alpha} + g^{\mu \nu} \mathcal{L}_{EM},
\end{equation}
up to a total derivative. We know from (\ref{boxh}) that, to first approximation, $T_{\mu \nu}$ is locally conserved, since
\begin{equation}
    \partial_\mu T^{\mu \nu} = - \frac{1}{16 \pi G} \Box \partial_\mu h^{\mu \nu} = 0,
\end{equation}
by virtue of the transverse-traceless condition\footnote{This is of course only the lowest order approximation to the true conservation law $\nabla_\mu T^{\mu \nu} = 0 $.}. One can also see this directly from (\ref{Tmn}):
\begin{equation}
    4 \pi \partial_\mu T^{\mu \nu} = (\partial_\mu \mathcal{G}^{\mu \alpha}) F^\nu_\alpha + \mathcal{G}^{\mu \alpha} (\partial_\mu F^\nu_\alpha) + g^{\mu \nu} \partial_\mu \mathcal{L}_{EM}.
\end{equation}
The first term is zero because of the equation of motion (\ref{Gtilde}). After contraction with the antisymmetric tensor $\mathcal{G}^{\mu \alpha}$, only the antisymmetric part of $\partial_\mu F^\nu_\alpha$ for the indices $(\mu, \alpha)$ survives in the second term. The third term can be worked out straightforwardly using the chain rule and the definition (\ref{Gdef}). Therefore,
\begin{equation}
    4 \pi \partial_\mu T^{\mu \nu} = \frac{1}{2} \mathcal{G}_{\mu \alpha} \left( \partial^\mu F^{\nu \alpha} + \partial^\alpha F^{\mu \nu} + \partial^\nu F^{\alpha \mu}\right) = 0,
\end{equation}
since the quantity in brackets is the Bianchi identity (\ref{Ftilde}).

Given (\ref{DH}), we can write the spatial components of the tensor (\ref{Tmn}) in terms of the physical fields as
\begin{equation}\label{Tij}
    4 \pi T_{ij} = - (D_i E_j + B_i H_j) + \frac{1}{2} \delta_{ij} (H \cdot B + D \cdot E),
\end{equation}
since $\mathcal{L}_{EM} = (\varepsilon E^2 - \mu^{-1} B^2)/2 \equiv (D\cdot E - H \cdot B)/2$. In our setup, $h_{12}$ is the only metric component that mixes linearly with the electromagnetic wave. Given the electromagnetic stress-energy tensor (\ref{Tij}), the gravitational wave equation (\ref{boxh}) for $h_{12}$ is
\begin{equation}\label{hdiff}
    \Ddot{h}_{12} - h''_{12} = -4 G B_0 \varepsilon v^2 b.
\end{equation}

Note the additional factors of $v^2$ and $\varepsilon$ in (\ref{bdiff}) and (\ref{hdiff}) compared to the wave equations of \cite{Pal} (see in particular equations 2.13 and 2.14 of the paper). At low energies $B_0 \ll B_c$, $\mu \approx 1$, $\varepsilon \approx 1$, and $v \approx 1$ (the speed of light), and the results of \cite{Pal} are retrieved.

The two coupled differential equations that describe graviton-photon oscillations in a polarized medium are 
\begin{align}\label{coupledbh}
    \begin{split}
        \Ddot{b} - v^2 b'' &= v^2 h'' B_0, \\
        \Ddot{h} - h'' &= -4 G B_0 \varepsilon v^2 b,
    \end{split}
\end{align}
where we write $h \equiv h_{12}$ to avoid clutter. The exact solution of (\ref{coupledbh}) can be written as
\begin{align}\label{ansatz}
    \begin{split}
        h &= \mathcal{A} e^{i (k z - \omega_g t)} + \frac{4 G B_0 \varepsilon v^2}{\omega_\gamma^2-k^2} \mathcal{B} e^{i (k z - \omega_\gamma t)},\\
        b &= \frac{\omega_g^2-k^2}{4 G B_0 \varepsilon v^2} \mathcal{A} e^{i (k z - \omega_g t)} + \mathcal{B} e^{i (k z - \omega_\gamma t)},
    \end{split}
\end{align}
where
\begin{equation}\label{wpm}
    \omega_{g,\gamma}(k) = \sqrt{\frac{1+v^2}{2} k^2 \pm \frac{1}{2}\sqrt{16 G B_0^2 \varepsilon v^4 k^2 + k^4(1-v^2)^2}}
\end{equation}
is the frequency of the graviton (photon) oscillation mode, and $k$ the wave number. Note first that for $B_0 \rightarrow 0$ the solution decouples into a sum of non-interacting fields with different group velocities, namely $\omega_g(k) \rightarrow k$, $\omega_\gamma(k) \rightarrow vk$, and $h \rightarrow \mathcal{A} e^{i k(z - t)}$, $b \rightarrow \mathcal{B} e^{i k(z - v t)}$, corresponding to freely propagating gravitational and electromagnetic fields, respectively.

For $B_0 \neq 0$, one can expand (\ref{wpm}) around $\omega_g = k$ or $\omega_\gamma = vk$ to extract the low frequency component $\omega_s$ of each oscillation mode:
\begin{align}\label{wggamma}
\begin{split}
    \omega_g(k) &\approx k - \frac{k(1-v)}{2} + \frac{1}{2} \sqrt{4 G B_0^2 \varepsilon v^4 + k^2(1-v)^2} \equiv k + \omega_s,\\
     \omega_\gamma(k) &\approx v k + \frac{k(1-v)}{2} - \frac{1}{2} \sqrt{4 G B_0^2 \varepsilon v^4 + k^2(1-v)^2} \equiv v k - \omega_s,
\end{split}
\end{align}
where it is assumed that $k \gg \omega_s$. In Dyson's wave packet formalism, the low frequency component describes the slow oscillation of the wave packet between gravitational and electromagnetic states \cite{Pal}. The requirement that $k \gg \omega_s$ is known as the slowly varying envelope approximation, which is the assumption that the envelope of the wave varies slowly compared to its characteristic frequency. We will use the same approximation in the next section to derive the quantum version of the Gertsenshtein effect.

Finally, we define the mixing frequency as the phase difference between the two oscillation modes, i.e. 
\begin{equation}\label{ws}
    \omega_m \equiv \frac{\omega_g - \omega_\gamma}{2} = \frac{1}{2} \sqrt{4 G B_0^2 \varepsilon v^4 + k^2(1-v)^2}.
\end{equation}
This concludes the derivation of the classical Gertsenshtein effect in a polarized vacuum. In the next section we show that the background magnetic field can similarly catalyze a resonant mixing between quantum states.

\section{Graviton-photon mixing matrix}\label{mixing}
The derivation of the Gertsenshtein effect presented in the previous section is (semi-)classical, and demonstrates oscillations between classical fields. Viewed as a quantum process, the same effect gives rise to graviton-photon mixing.

Defining the two component field $\psi = (b, h)^T$, and working in the slowly varying envelope approximation, one can recast (\ref{coupledbh}) as the following first order differential equation \cite{Ejlli}
\begin{equation}\label{schrodinger}
    i \partial_t \psi = \mathcal{H} \psi,
\end{equation}
where
\begin{equation}\label{H}
    \mathcal{H} = \begin{pmatrix}
vk & i \sqrt{\varepsilon G} B_0 v^2\\
-i \sqrt{\varepsilon G} B_0 v^2 & k
\end{pmatrix}
\end{equation}
is the Hamiltonian of the combined system. Note that the eigenvalues of (\ref{H}) are $\omega_g$ and $\omega_\gamma$ defined in the previous section. The general solution of (\ref{schrodinger}) is
\begin{equation}\label{ket}
    \ket{\psi} = e^{-i \mathcal{H} t} \ket{\psi}_0 \equiv \mathcal{M} \ket{\psi}_0,
\end{equation}
where we define $\mathcal{M} \equiv e^{-iHt}$ as the (manifestly unitary) mixing matrix, and we write the mixed quantum state $\ket{\psi}$ as a sum over the basis states $\ket{\psi} = b_0 \ket{b} + h_0 \ket{h}$, where $|b_0|^2+|h_0|^2 = 1$, and $\ket{b} = (1,0)^T$, $\ket{h} = (0,1)^T$ represent a purely electromagnetic and a purely gravitational state, respectively.

The complex exponential of a 2x2 matrix is given by \cite{Domcke}
\begin{equation}
    e^{-i \mathcal{H} t} = e^{-i/2 \Tr(\mathcal{H}) t} \left[ \cos\left(\frac{\omega_g-\omega_\gamma}{2}t\right) \mathds{1} - \frac{2i}{\omega_g - \omega_\gamma} \sin\left(\frac{\omega_g-\omega_\gamma}{2}t\right) \left(\mathcal{H}-\frac{1}{2} \Tr(\mathcal{H})  \mathds{1} \right) \right],
\end{equation}
therefore
\begin{equation}
    \mathcal{M} = e^{-ik(1+v)t/2}\begin{pmatrix}
\cos \theta + i \sqrt{1-f^2} \sin \theta & f \sin \theta \\
    -f \sin \theta & \cos \theta - i \sqrt{1-f^2} \sin \theta
\end{pmatrix},
\end{equation}
where $\theta \equiv \omega_m t$ is the mixing angle and $f \equiv \sqrt{\varepsilon G} B_0 v^2/\omega_m$ the suppression factor responsible for partial mixing. In the decoupled limit $B_0\rightarrow 0$, $\theta \rightarrow k(1-v)/2$, $f \rightarrow 0$ and $\mathcal{M} \rightarrow \text{diag}(e^{-i v k}, e^{-ik})$, while in the unpolarized limit $v \rightarrow 1$, $f \rightarrow 1$ and $\mathcal{M}$ is the usual rotational mixing matrix.

Assuming that the initial state is purely gravitational, $\ket{\psi}_0 = \ket{h}$, after a distance $t=D$ the mixed state will be 
\begin{equation}
    \ket{\psi} =e^{-ik(1+v)D/2} \begin{pmatrix}
           f \sin\left(\omega_m D\right)  \\
           \cos \left(\omega_m D\right) - i \sqrt{1-f^2} \sin \left(\omega_m D\right)
         \end{pmatrix}.
\end{equation}
Therefore, the probability of graviton-to-photon conversion is
\begin{equation}\label{prob}
    \mathcal{P}(g \rightarrow p) \equiv |\bra{b}\ket{\psi}|^2 = f^2 \sin^2\left(\omega_m D\right) = \frac{G B_0^2 \varepsilon v^4}{\omega_m^2}\sin^2\left(\omega_m D\right).
\end{equation}
Using (\ref{ws}), this is 
\begin{equation}\label{prob2}
    \mathcal{P}(g \rightarrow p) = \frac{4 G B_0^2 \varepsilon v^4}{4 G B_0^2 \varepsilon v^4 + k^2(1-v)^2} \sin^2\left(\frac{1}{2} \sqrt{4GB_0^2 \varepsilon v^4 + k^2(1-v)^2} D\right).
\end{equation}
Given that $0 \leq v \leq 1$, $\mathcal{P} \leq 1$. Whenever $v \neq 1$, the probability is strictly less than one. This indicates that when vacuum polarization effects are taken into account the conversion of gravity into light (and viceversa) is only partial. In the decoupled limit ($B_0 \rightarrow 0$) there is no mixing and the probability vanishes.

The results of this section and the previous are completely general and valid for any field strength. In order to proceed further, we need explicit expressions for the electric permittivity and the magnetic permeability of the vacuum in (\ref{emu}), which we derive in the next two sections in both the weak and strong field limit.

\section{Graviton-photon oscillations in the weak field limit}\label{weak}
In the weak field limit $E \ll E_c$ and $B \ll B_c$, therefore $a,b \ll 1$. Additionally, in our setup the magnetic field dominates over the other components, therefore $b \gg a$.

In this limit one can expand the trigonometric functions $x\cot x$ and $x\coth x$ in (\ref{oneloop}) in a power series around $x=0$ (see appendix \ref{remarks} for details). Keeping only the lowest order terms, one can write
\begin{equation}\label{cotcoth}
    (as)(bs) \cot(as) \coth(bs)  \approx 1 + \frac{s^2}{3} (b^2-a^2) - s^4 \left( \frac{b^4}{45} + \frac{a^4}{45} + \frac{a^2b^2}{9} \right).
\end{equation}
The first two terms cancel the last two of (\ref{oneloop}), and one is left with
\begin{equation}\label{weakL}
    \mathcal{L}_{EM}^{(1)}(a,b) \approx \frac{m_e^4}{8 \pi^2} \int_0^\infty ds \, s \, e^{-s} \left( \frac{b^4}{45} + \frac{a^4}{45} + \frac{a^2b^2}{9} \right) = \frac{m_e^4}{360 \pi^2} (a^4+b^4+5a^2b^2).
\end{equation}
Given the definitions in (\ref{ab}), this is nothing but the Euler-Heisenberg Lagrangian \cite{EH}
\begin{equation}\label{EH}
    \mathcal{L}_{EH} = \frac{q^4}{360 \pi^2} \left(4 S^2 + 7P^2 \right) = \frac{\alpha^2}{90 m_e^4} \left[ (F_{\mu \nu}F^{\mu \nu})^2 + \frac{7}{4} (\Tilde{F}_{\mu \nu}F^{\mu \nu})^2 \right].
\end{equation}

Having the closed form (\ref{weakL}) for the electromagnetic Lagrangian as a function of $a$ and $b$ allows us to compute the electric permittivity and magnetic permeability in (\ref{emu}). In the weak field limit, and for $b \gg a$, these are
\begin{align}\label{DEHB}
    \begin{split}
        \varepsilon &= 1 + \frac{q^2 b^2}{36 \pi^2} = 1 + \frac{4 \alpha^2 B_0^2}{9 m_e^4},\\
        \frac{1}{\mu} &=  1 - \frac{q^2b^2}{90 \pi^2}= 1 - \frac{8 \alpha^2 B_0^2}{45 m_e^4}.
    \end{split}
\end{align}
The group velocity of the electromagnetic radiation then is
\begin{equation}\label{v2weak}
    v^2 = \frac{1}{\varepsilon \mu} = \frac{1 - \frac{8 \alpha^2 B_0^2}{45 m_e^4}}{1 + \frac{4 \alpha^2 B_0^2}{9 m_e^4}}
\end{equation}
For $B_0 \ll B_c$, $v^2 \rightarrow 1$ and the mixing frequency (\ref{ws}) reduces to the known result $\omega_m = \sqrt{G}B_0$. It is also clear from (\ref{prob2}) and (\ref{v2weak}) that the mixing probability decreases with increasing $B_0$. Indeed, close to the Schwinger limit $v^2 \rightarrow 0$, and $\mathcal{P} \rightarrow 0$. Physically, this is due to the fact that the non-linear interactions encoded in (\ref{EH}) slow down the electromagnetic radiation, but not gravity, which always propagates at the speed of light in vacuum. Assuming they start in sync, after a distance $\Delta x$ the phase difference between the two waves will be $\Delta \phi = k (1-v) \Delta x$, which reaches order one at the decoherence length \cite{Dyson}
\begin{equation}\label{Lc}
    L_c = \frac{1}{k (1-v)} \approx \frac{45 m_e^4}{14 \alpha^2 B_0^2 k}.
\end{equation}
For values of the background magnetic field small compared to the critical value $B_c$, $v \approx 1$ and $L_c > 2/\sqrt{G}B_0$, therefore decoherence takes place after mixing and one retrieves the classical result. On the other hand, when $B_0$ reaches values close to the Schwinger limit, the decoherence length becomes smaller than the (classical) mixing length, and prevents oscillations above that scale. Beyond the Schwinger limit the Euler-Heisenberg Lagrangian is no longer sufficient, and one needs to consider higher order terms in (\ref{cotcoth}). However, as we show in \S \ref{strong}, it is reasonable to expect that beyond this limit oscillations cease altogether.

\section{Graviton-photon oscillations in the strong field limit} \label{strong}
In our setup, the strong field regime corresponds to $B \gg B_c$, but $E \ll E_c$, therefore the relevant limit is $a \ll 1$ and $b \gg 1$. As in the previous section, we need to compute the derivatives of (\ref{oneloop}) with respect to $a$ and $b$, for $a \ll 1$ and $b \gg 1$\footnote{Note that it is inconsistent to first set $a=0$ in (\ref{oneloop}) and then compute the derivatives, as this procedure always leads to $\Vec{\Pi} = 0$. I would like to thank Mikhail Medvedev for clarifying this point.}. 

Integration of the individual terms in (\ref{oneloop}) leads to divergences. However, if one regularizes the integrals, the divergent pieces cancel out, leaving only the finite part. The calculation is involved and is presented in details in appendix \ref{appendix}. The integrals are \cite{Dittrich, Lundin, Kim, Medvedev}
\begin{align}
    \begin{split}
        \frac{\partial \mathcal{L}_{EM}^{(1)}(a,b)}{\partial a} &= \frac{a m_e^4}{12 \pi^2} \left( b - \ln b \right), \\
        \frac{\partial \mathcal{L}_{EM}^{(1)}(a,b)}{\partial b} &= \frac{b m_e^4}{12 \pi^2} \ln b.
    \end{split}
\end{align}
The electric permittivity and magnetic permeability are then given by (\ref{emu}):
\begin{align}
    \begin{split}
        \varepsilon =  1 + \frac{\alpha}{3 \pi}\left( b - \ln b \right), \\
        \mu^{-1} = 1 - \frac{\alpha}{3 \pi} \ln b.
    \end{split}
\end{align}
Therefore, the group velocity is
\begin{equation}
    v^2 = \frac{1}{\varepsilon \mu} \approx \frac{1 - \frac{\alpha}{3\pi} \ln \left( \frac{B_0}{B_c}\right)}{1 + \frac{\alpha}{3 \pi} \frac{B_0}{B_c}}.
\end{equation}
In the strong field limit $B_0 \gg B_c$, and $v \approx 0$. When the group velocity is zero, the electromagnetic field effectively stops propagating, and the mixing probability (\ref{prob2}) vanishes, vindicating our expectation that oscillations should stop above the Schwinger limit.

\section{The Gertsenshtein detector} \label{experiment}
Given what we know about the Gertsenshtein effect in the non-linear electromagnetic regime, is it possible, in principle, to detect a single graviton using graviton-to-photon conversion? 

One can set up a basic experiment that would be able to achieve this by filling a region of space of linear dimension $D$ with a background magnetic field of intensity $B_0$. If a graviton of energy $k$ is incident upon this region, it is converted to a photon with probability (\ref{prob}):
\begin{equation}
    \mathcal{P} = \frac{G B_0^2 \varepsilon v^4}{\omega_m^2}\sin^2\left(\omega_m D\right),
\end{equation}
where the mixing frequency $\omega_m$ is given by (\ref{ws}). The photon can then be detected with very high probability by a photodetector and the quantization of the gravitational field inferred indirectly. The probability of conversion above thus measures the efficiency of the detector as a whole. We will say that the detector is efficient at detecting gravitons if, across a distance $D$, $\mathcal{P} \sim \mathcal{O}(1)$.

If we choose $D$ of the order of the oscillation length, $D \sim \omega_m^{-1}$, to maximize the probability of detection, the condition for an efficient detector is
\begin{equation}\label{condition}
    \frac{4 G B_0^2 \varepsilon v^4}{4 G B_0^2 \varepsilon v^4 + k^2(1-v)^2}\sim \mathcal{O}(1).
\end{equation}
The crucial question is: can this be achieved in our universe?

We know from \S \ref{strong} that oscillations stop above the Schwinger limit, so we can restrict our analysis to the weak field regime $B_0 < B_c$. The probability above approaches 1 in the decoupled limit $B_0 \rightarrow 0$. In this limit, however, the mixing length becomes infinite, so our goal is to minimize the mixing length as much as possible while maintaining a sizeable probability of detection. The question then becomes: what is the maximum value of $B_0$ before the probability of detection is no longer appreciable? Or, in other words, what is the minimum size of an efficient detector?

From (\ref{condition}) it is clear that the condition is satisfied whenever $4GB_0^2 \gtrsim k^2(1-v)^2$. Given that $B_0 \ll B_c$, $1-v \approx 14 \alpha^2 B_0^2/45 m_e^4$, so $B_0 \lesssim \sqrt{G} m_e^4/\alpha^2 k$. Thus the maximum possible value of the magnetic field before coherence is destroyed is\footnote{At this magnetic field strength, the probability of detection is around $0.5$.}
\begin{equation}\label{B0bar}
    \Bar{B}_0 \sim \frac{\sqrt{G} m_e^4}{\alpha^2 k}.
\end{equation}
One can arrive at the same condition by equating the coherence length (\ref{Lc}) with the classical mixing length: $k(1-v) = \sqrt{G} \bar{B}_0/2$. Note also that, since we are assuming $\Bar{B}_0 \ll B_c$, we need $k \gg \sqrt{G} B_c/\alpha$. 

The  minimum size $\bar{D}$ for an efficient detector is the mixing length corresponding to (\ref{B0bar}), namely
\begin{equation}\label{D}
    \bar{D} \sim \omega_m^{-1} = \frac{2}{\sqrt{G} \Bar{B}_0} \sim \frac{\alpha^2 k}{G m_e^4}.
\end{equation}
 For an X-ray photodetector like the one used by the CERN Axion Solar Telescope (CAST)\footnote{CAST is currently used to detect axions via axion-photon mixing. The same type of device could be used to detect single gravitons via graviton-photon mixing \cite{Carney}.}\cite{CAST}, $k \sim 10^{18} \, \text{Hz} \sim 10 \, \text{keV}$, and $\Bar{D} \sim 10^{23} \, \text{km}$, approximately the radius of the observable universe.

In principle, one could consider graviton frequencies much smaller than $10^{18}$ Hz, as the only theoretical constraint on $k$ is $k \gg \sqrt{G} B_c/\alpha$, which is easily satisfied. For example, if one takes $k \sim$ kHz, $\Bar{D} \sim 10^8$ km, while the graviton/photon wavelength is $k^{-1} \sim 300 \, \text{km} \ll \Bar{D}$. However, in such a setup, photon detection becomes infeasible. Modern low-frequency single-photon detectors are sensitive to near infrared wavelengths of up to 1 $\mu$m \cite{Natarajan,Bai}, many orders of magnitude below the km range needed to (realistically) perform the experiment. Photon wavelengths in the micrometer to nanometer range inevitably give cosmological values for the detector size, roughly in the range from Mpc to Gpc.

At cosmological scales, however, one should take into account the effects an expanding universe has on distances and fields \cite{Anninos}. Given that the magnetic field scales as $a^{-2}$, the proper mixing length is $D_p = 2a^2/\sqrt{G} B_0$, while the Hubble radius is $H^{-1} = a/\dot{a}$. Therefore, in general 
\begin{equation}\label{HDp}
    \frac{D_p}{H^{-1}} = \frac{2 a \dot{a}}{\sqrt{G} B_0}.
\end{equation}
Using Friedmann's equation $H^2 = (8 \pi G/3) \rho$, one can rewrite (\ref{HDp}) as
\begin{equation}
     \frac{D_p}{H^{-1}} = \sqrt{\frac{32 \pi \rho}{3}} \frac{a^2}{B_0}.
\end{equation}
For a perfect fluid with equation of state $p = \omega \rho$, $\rho \propto a^{-3(1+\omega)}$, therefore
\begin{equation}\label{ratio}
    \frac{D_p}{H^{-1}} = \sqrt{\frac{32 \pi \rho_0}{3}} \frac{a^\frac{{1-3\omega}}{2}}{B_0}.
\end{equation}
For $\omega < 1/3$ (which includes the matter and dark energy dominated cases), the ratio in (\ref{ratio}) increases with the expansion, exponentially in deSitter space. In the current stage of the universe's evolution, this means that the proper mixing length very quickly becomes larger than the horizon. Furthermore, assuming the intergalactic magnetic field does not dominate the energy density of the universe, $B_0^2 \lesssim \rho_0$, (\ref{ratio}) entails
\begin{equation}
    D_p \gtrsim \sqrt{\frac{32 \pi}{3}} H^{-1},
\end{equation}
larger than the Hubble radius\footnote{Our analysis is one of principle, and as such we do not wish to discuss in details the practical limitations of building a graviton-photon mixing detector in the real universe. However, for detectors of cosmological size, one should at a minimum note that in the standard cosmological model the Gertsenshtein mechanism is inoperative above recombination temperatures due to Thomson scattering, which makes the universe opaque to electromagnetic radiation \cite{Anninos}. After recombination but before reionization (the ``dark ages'') the presence of charged particles further lowers the speed of electromagnetic radiation \cite{Domcke}, making detection even more difficult. This only strengthen our conclusion that graviton detection by the Gertsenshtein effect is exceedingly unlikely in the real world.}. The conclusion is that when one factors in the expansion of the universe, graviton-photon oscillations on cosmological scales always lie outside the Hubble radius and are as a result unobservable. 

Therefore, barring any futuristic single photon detection technology that would allow for detection frequencies much smaller than a THz, the detection of a single graviton using the Gertsenshtein mechanism is generically inefficient (in any remotely realistic scenario) due to nonlinear vacuum polarization effects. Note the crucial role played by the electromagnetic constants in our conclusion: if the fine structure constant were much smaller, or the electron mass much larger, than the values they take in our universe, the Gertsenshtein experiment could in principle be performed successfully on astrophysical or even planetary scales\footnote{Choosing different values for the electromagnetic constants does not necessarily impact the efficiency of photon detection. Using a simple toy model for photon detection consisting of a gas of hydrogen atoms, their binding energy is $E_B = -m_e \alpha^2/2$. One could increase the electron mass while at the same time decreasing the fine structure constant in such a way that the binding energy stays the same, and the photoemission process happens at the same frequency. Thus, it is possible to imagine a universe in which (single) graviton-photon oscillations are observable \textit{and} photon detection is efficient at X-rays frequencies.}. 

Our conclusion of course only applies to the detection of a \textit{single} graviton: the number of photons detected in the experiment is roughly equal to $N_\gamma \sim \mathcal{P} N_G$, where $N_G$ is the number of gravitons incident upon the magnetic region. Even when $\mathcal{P} \ll 1$, $N_\gamma$ can be made greater than one if $N_G$ is large enough, and for relatively small detector sizes. When $N_G \gg 1$, however, we are no longer probing the quantum gravitational regime: we simply have a classical gravitational wave. 

In the Gertsenshtein experiment, the nonlinearities of the electromagnetic field seem to conspire to hide the quantization of gravity. Interestingly, the obstruction to probing the quantum gravitational regime in the Gertsenshtein experiment is then qualitatively different than the one encountered in other proposed graviton detector architectures, which we explore in the next section.

\section{Relation to other experiments} \label{relation}
Many different graviton detector architectures have been considered in the literature \cite{Dyson, Rothman, Boughn, Krauss}, and, so far, all of them have been shown to be inefficient. In this section, we revisit them, trying to find the common reason behind their inefficiency, and compare them with the Gertsenshtein experiment of \S \ref{experiment}.\\

The simplest kind of graviton detector is a collection of atoms. Dyson showed that an atom absorbs a graviton with a planckian cross section:
\begin{equation}
    \sigma_{\text{abs}} = 4 \pi^2 l_p^2 Q,
\end{equation}
where $l_p$ is the Planck length and $Q$ is a number of order one\footnote{In fact, this result extends to generic (quantum) bound states \cite{Pal2}.}. A collection of atoms is efficient at absorbing gravitons if the optical depth of the graviton through the medium is greater than one:
\begin{equation}\label{opticaldepth}
    n_B \sigma_{\text{abs}} D > 1,
\end{equation}
where $n_B$ is the density of atoms (bound states) and $D$ the size of the detector. Assuming the mass of the detector is $M$, while the mass of a single atom is $m_B$, then $n_B = M/(m_B D^3)$, and the optical depth is
\begin{equation}
    n_B \sigma_{\text{abs}} D \sim \frac{G M}{D} \frac{1}{m_B D} \lesssim 1,
\end{equation}
given that $GM < D$ to avoid gravitational collapse and $m_B D > 1$ for particles whose Compton wavelength is smaller than the detector. Therefore graviton absorption is generically inefficient. Barring any high-energy modification of General Relativity, this also prevents the occurrence of gravitational absorption lines in the primordial gravitational wave spectrum \cite{Pal2}.\\

\noindent Architectures based on laser interferometry, such as the one used in LIGO, fare no better. The energy density of a gravitational wave of strain $f$ and frequency $\omega$ is
\begin{equation}
    \rho = \frac{f^2 \omega^2}{32 \pi G},
\end{equation}
while the energy density of a single graviton of the same frequency is $\omega^4$. Therefore, in order to detect a single graviton, a LIGO-type apparatus needs to be sensitive to a strain
\begin{equation}
    f = \sqrt{32 \pi} \omega l_p.
\end{equation}
The measured strain is nothing but the fractional change in the distance $D$ between two masses: $f = \Delta D/D$. Thus the detector  needs to be sensitive to a distance variation of 
\begin{equation}
    \Delta D = f D = \sqrt{32 \pi} (\omega D) l_p \lesssim \sqrt{32 \pi} l_p,
\end{equation}
given that the distance $D$ between the two masses has to be smaller than the graviton wavelength: $D < \omega^{-1}$. Consequently, a LIGO-type interferometer needs to measure distances to Planck length accuracy in order to be able to detect a single graviton, and is therefore technically inefficient at this task.\\

\noindent A kind of cosmic detector was envisioned in \cite{Krauss}. Here, the authors point out that the observation of a cosmological gravitational wave background associated with an inflationary phase would provide evidence for the quantization of the gravitational field. However, we know that if the inflationary phase lasts too long, the length scales we observe nowadays correspond to modes smaller than the Planck length at the onset of inflation. This is the trans-Planckian problem of inflationary cosmology \cite{Martin, Niemeyer}. It has been conjectured \cite{TPCC} that this problem can never arise in any consistent UV-complete theory of gravity. According to this trans-Planckian Censorship Conjecture (TCC), no mode that exits the Hubble horizon could ever have had a wavelength smaller than the Planck length. If $a_i$ and $a_f$ are the values of the scale factor at the beginning and end of the inflation phase, respectively, the TCC reads \cite{Bedroya}
\begin{equation}\label{TCC}
    \frac{a_f}{a_i} < \frac{m_p}{H_f},
\end{equation}
where $H_f$ is the Hubble rate at the end of inflation. Given that after inflation $a_f/a_i$ is exponentially large, the condition above severely limits the scale of inflation $H_f$. In particular, if inflation provides a causal mechanism for the origin of structure, the current Hubble volume must originate from the Hubble volume at the beginning of inflation, meaning that roughly
\begin{equation}\label{H0}
    \frac{1}{H_0} \approx \frac{1}{H_f} e^N \frac{T_R}{T_0},
\end{equation}
where $H_0$ is the Hubble rate today, $N$ the number of e-foldings, and $T_0$, $T_R$ the temperatures today and at reheating, respectively. If reheating is fast, $T_R \approx \sqrt{m_p H_f}$. Furthermore, given that the matter energy density today is larger than the radiation energy density by a factor $T_{eq}/T_0$, where $T_{eq}$ is the temperature at matter-radiation equality, $H_0 = \sqrt{T_0^3 T_{eq}}/m_p$, and (\ref{H0}) gives
\begin{equation}
    \frac{a_f}{a_i} = e^N \sim \sqrt{\frac{m_p H_f}{T_0 T_{eq}}}.
\end{equation}
Plugging this back into (\ref{TCC}) gives the condition
\begin{equation}
    H_f < (m_p T_0 T_{eq})^{1/3},
\end{equation}
forcing the tensor-to-scalar ratio $r$ to be 
\begin{equation}
    r \sim 10^{8} \left(\frac{H_f}{m_p} \right)^2 < 10^{-30},
\end{equation}
practically unobservable \cite{Bedroya}. In short, if one makes the minimal assumption that subplanckian modes during inflation cannot cross the horizon, classicalize, and become observable (TCC), tensor modes are also generically unobservable and cannot be used to establish the quantization of gravity. Any model of inflation that predicts observable tensor modes also makes predictions about physical quantities today that rely on unknown trans-Planckian physics, and its validity is therefore dubious.\\

\noindent In the first two experiments just reviewed, the obstruction in observing the quantization of the gravitational field seems to stem from the impossibility to resolve scales below the Planck length. This fundamental limit in the precision of distance measurements is itself a consequence of gravitational collapse in General Relativity \cite{Calmet}: trying to resolve finer distances requires so much energy that the system is bound to collapse to a black hole. In a sense then, the parameter space that would enable graviton detection is always hidden behind the event horizon of a black hole. It is important to realize, however, that this conclusion relies on the assumption that gravitational collapse generically results in black holes, i.e. singularities hidden behind event horizons. This is the content of the (weak) Cosmic Censorship Conjecture (CCC) \cite{CCC, Penrose}. All we can say, then, is that in a universe in which the CCC is true, graviton detection architectures based on laser interferometry or graviton absorption are generically inefficient. Interestingly, in the third experiment we reviewed it is the TCC that, if true, would impede graviton detection. In this scenario, in place of a black hole horizon, fluctuations in the trans-Planckian regime are instead hidden behind the cosmological horizon so as to make them unable to turn classical and affect macroscopic observations \cite{Brahma}.

Therefore, in the graviton detector architectures discussed above the nonlinear gravitational physics responsible for the formation of an event horizon conspires to hide the quantization of the gravitational field. Note then that the Gertsenshtein experiment discussed in \S \ref{experiment} is qualitatively different in that the electromagnetic field plays the role of the co-conspirator: the nonlinearities of the electromagnetic field destroy the coherence of the two waves to prevent oscillations at scales below the Hubble radius. Vacuum polarization is instrumental in securing inefficiency: in its absence, oscillations would occur inside the cosmological horizon and become observable.

In fact, if one were to ignore vacuum polarization effects (equivalent to taking $\alpha = 0$ or $m_e = m_p$) the probability of conversion would simply be
\begin{equation}
    \mathcal{P} = \sin^2(\omega_m D),
\end{equation}
with $\omega_m = \sqrt{G} B_0$. Fixing $\mathcal{P}=1$, one could then make $D$ arbitrarily small by increasing the strength of the magnetic field $B_0$ beyond the Schwinger limit. Graviton-to-photon conversion would happen with probability 1 at arbitrarily small scales, independently of the frequency of the incoming graviton. In such a universe, single graviton detection using the Gertsenshtein effect would in principle be possible in the laboratory.

\section{Conclusions}
It is often assumed that gravity is quantized. Given that the quantization of gravity does not follow logically from the quantization of matter \cite{Bohr}, its existence can only be experimentally decided. The Cosmic Censorship Conjecture \cite{CCC} entails that quantum gravity effects are always hidden behind event horizons. A particular form of cosmic censorship comes into play in the detection of single gravitons: Dyson, Rothman and Boughn \cite{Dyson, Rothman} proposed several graviton detector architectures and showed they cannot work due to gravitational collapse.

In this paper, we have studied a graviton detector architecture based on graviton-to-photon conversion (originally proposed by Dyson) and we have shown that, for photon frequencies $k$ consistent with current single-photon detector sensitivities, this architecture is generically inefficient at detecting gravitons. This is due to the nonlinearities of the electromagnetic field, which destroy graviton-photon coherence above the scale $\Bar{B}_0 \sim \sqrt{G} m_e^4/\alpha^2 k$. For typical single-photon detector frequencies, the magnetic field strength is such that graviton-photon oscillations occur at cosmological scales, and one needs to take into account the expansion of the universe. A consistent treatment of graviton-photon oscillations in the cosmological setting reveals that these always lie outside the Hubble radius and are in principle unobservable \cite{Anninos}.

Contrary to other graviton detectors discussed in \S \ref{relation}, where the inefficiency is purely a consequence of the nonlinear gravitational processes that lead to the formation of an event horizon, the inefficiency of the Gertsenshtein detector is partly a result of the nonlinearity of the electromagnetic field in the form of vacuum polarization effects: in a hypothetical scenario with no vacuum polarization, oscillations would occur well within the cosmological horizon, making single graviton detection possible. However, while gravitational collapse constitutes a clear-cut obstruction to graviton detection, the coherence breaking effect of vacuum polarization makes graviton detection practically, but not theoretically, impossible. In particular, a futuristic technology capable of detecting single photons at wavelengths much larger than the nanometer range of currently existing detectors could in principle be envisioned. If such frequencies could be achieved, an efficient detector would be of astrophysical, instead of cosmological, size.

The nonlinearities of the gravitational field cloak quantum gravitational effects behind event horizons. With the caveats discussed above, in the Gertsenshtein experiment the nonlinearities of the electromagnetic field force oscillations to lie beyond the cosmological horizon, so that graviton detection is just outside the reach of empirical observation. If one assumes that the inability to detect a single graviton has a deeper origin, for example if the gravitational field arises as a purely classical entity in the thermodynamic limit of some microscopic non-gravitational theory \cite{Jacobson, Padmanabhan, Verlinde}, the different obstructions in the two cases could be suggestive of a duality between electromagnetism and gravity, in the spirit of \cite{AdS/CFT}. This remains the subject for future investigation.

\section*{Acknowledgement}
I would like to thank Tony Rothman for providing invaluable feedback on an early version of the draft, which greatly helped me clarify the presentation.

\appendix

\section{Remarks on the one-loop effective Lagrangian}\label{remarks}
In this appendix we discuss some mathematical subtleties related to the 1-loop effective QED Lagrangian (\ref{oneloop}), and justify our choice of disregarding a similar contribution due to gravity.

Since the expression in (\ref{oneloop}) has first-order poles at $s = n \pi/a$ with $n=1,2,...$, the integral representation must be defined with an appropriate regularization scheme \cite{Jentschura}:
\begin{equation}\label{regL}
    \mathcal{L}_{EM}^{(1)} = - \frac{m_e^4}{8 \pi^2} \lim_{\epsilon, \eta \rightarrow 0^+} \int_{i \eta}^{\infty+i\eta} ds \frac{e^{-s(1-i\epsilon)}}{s^3} \left[ (as)(bs) \cot(as) \coth(bs) -1 +\frac{s^2}{3} (a^2-b^2)\right].
\end{equation}
Physically, the one-loop effective Lagrangian above has a natural perturbative expansion in (even) powers of the external photon field \cite{Dunne}:
\begin{equation}\label{feynseries}
    \includegraphics[width=0.8 \textwidth]{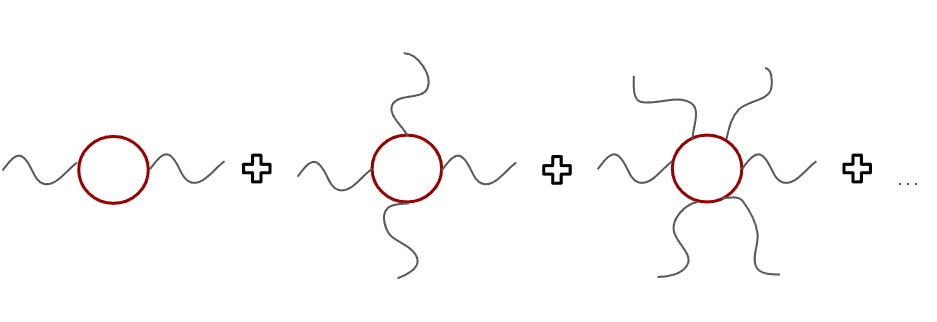}
\end{equation}
where the squiggly lines represent photons and the loops electron-positron pairs.

To see this, one can expand the trigonometric functions $x \coth x$ and $x \cot x$ around $x=0$:
\begin{equation}
    x \coth x = \sum_{n=0}^\infty \frac{2^{2n} \mathcal{B}_{2n}}{(2n)!}x^{2n}, \quad x \cot x = \sum_{m=0}^\infty \frac{(-1)^m 2^{2m} \mathcal{B}_{2m}}{(2m)!}x^{2m},
\end{equation}
where $\mathcal{B}$ are the Bernoulli numbers.
Substituting the power series into (\ref{regL}), one gets the weak field expansion
\begin{equation}\label{expansion}
   \mathcal{L}_{EM}^{(1)} \sim - \frac{m_e^4}{8 \pi^2} \sum_{n=2}^\infty\sum_{m=0}^n \frac{(2n-3)!}{(2m)!(2n-2m)!} \mathcal{B}_{2m} \mathcal{B}_{2n-2m} (2ia)^{2m} (2b)^{2n-2m}. 
\end{equation}
This is an expansion in the background electromagnetic field strength. For example, the first two orders of (\ref{expansion}) yield
\begin{equation}\label{expab}
    \mathcal{L}_{EM}^{(1)} \sim \frac{m_e^4}{360 \pi^2}(a^4 + 5 a^2b^2 + b^4) + \frac{m_e^4}{630 \pi^2} \left( a^6 + \frac{7}{2} a^4b^2 - \frac{7}{2} a^2b^4 -b^6\right) + ...
\end{equation}
The leading term encodes the correction to photon propagation due to light-light scattering and corresponds to the second diagram in (\ref{feynseries}), while the second term corresponds to the third diagram in (\ref{feynseries}) with six external photon legs and is further suppressed by additional factors of $B_c = m_e^2/q$. Note also that the first term of (\ref{expab}) coincides with (\ref{weakL}), as it should.
Similarly, it can be shown that the strong (magnetic) field expansion is
\begin{equation}\label{strongexp}
    \mathcal{L}_{EM}^{(1)} \sim \frac{a^2 m_e^4}{24 \pi^2}\left(- \psi\left(\frac{1}{2b}\right) - b - \ln(2b) \right) + \frac{b^2 m_e^4}{24 \pi^2} \left( \log 2b - \frac{1}{2}\right) + ...
\end{equation}

In general, one cannot exchange integral and infinite summation as we did above. The price to pay for this infraction is that (\ref{expab}) and (\ref{strongexp}) are \textit{asymptotic} expansions, namely formally divergent power series whose truncation at finite order can still provide a good approximation to the underlying function (in our case, the one-loop Lagrangian). In sections \ref{weak} and \ref{strong} we compute the electric permittivity and magnetic permeability of the vacuum by differentiating the one-loop Lagrangian with respect to the parameters $a$ and $b$ using (\ref{emu}). These derivatives should always be understood as computed on the corresponding (truncated) asymptotic series.  

In some cases, it is possible to restore the original non-perturbative information by resummation methods. Indeed, one can show that for a purely magnetic background, which is the case we are interested in, the series (\ref{expansion}) is diverging but alternating, and therefore Borel summable. One can also show that the Borel resummation of (\ref{expansion}) coincides with (\ref{regL}) \cite{Chadha}. For a purely electric background, on the other hand, the series is divergent and non-alternating, and thus is not even Borel summable, which signals the presence of additional non-perturbative sectors of the theory. Indeed, the Lagrangian (\ref{regL}) has an imaginary nonperturbative contribution which arises from the poles on the real $s$ axis and corresponds physically to vacuum instability due to electron-positron pair production in the presence of a static electric field. The full nonperturbative imaginary part of (\ref{regL}) was first derived by Schwinger \cite{Schwinger} and is given by
\begin{equation}
    \text{Im} \mathcal{L}_{EM}^{(1)} = \frac{q^2E^2}{8 \pi^3} \sum_{n=1}^\infty \exp\left( -\frac{\pi n E_c}{E}\right).
\end{equation}
In our setup, this contribution is vanishingly small as $E \ll E_c$.\\

The general structure of (\ref{expansion}) is that of a low energy effective field theory: the Lagrangian describes the physics of some light degrees of freedom (the photon field) by integrating out the heavy degrees of freedom (the electron field) that lie at or above a certain energy scale (the Schwinger limit). Schematically, given a heavy mass scale $m$ and a series of gauge and Lorentz invariant operators $O_{n}$ for the light fields of mass dimension $n$, one can write the effective Lagrangian as
\begin{equation}\label{Leff}
    \mathcal{L}_{\text{eff}} = m^4 \sum_n a_n \frac{O_{n}}{m^n},
\end{equation}
where $a_n$ are numerical coefficients. For example, the first nontrivial term of (\ref{expansion}) is the Euler-Heisenberg Lagrangian, with mass dimension $n=8$ and $O_8 = (F_{\mu \nu}F^{\mu \nu})^2 + 7/4 (\Tilde{F}_{\mu \nu} F^{\mu \nu})^2$. The next term in the expansion has mass dimension $n=12$, as one can see from (\ref{expab}). 

The one-loop effective Lagrangian for gravity has the same structure as (\ref{Leff}), just with different operators and, presumably, mass scale. If we assume, as we do in the main text, that the UV completion of gravity lies at the Planck scale, then $m \equiv m_p$ and the higher order contributions in the effective Lagrangian for gravity are completely negligible. In certain models, for example in models with large extra dimensions, $m$ can be much smaller and the loop corrections can become significant. 

The general structure of the effective field theory of gravity coupled to the Standard Model was analyzed in \cite{Ruhdorfer}. The first non-trivial effective coupling between gravity and electromagnetism arises at sixth order and has the form
\begin{equation}\label{L6}
    \mathcal{L}_{6} = \frac{a_6}{m_p^2} F^{\mu \nu} F^{\rho \sigma} C_{\mu \nu \rho \sigma},
\end{equation}
where $C_{\mu \nu \rho \sigma}$ is the Weyl tensor and $a_6$ a numerical coefficient of order 1. Schematically, this term is $\mathcal{L}_6 \sim m_p^{-2} F^2 \partial h \partial h$. By (\ref{coupled}), (\ref{L6}) gives a contribution to the graviton propagator $\sim \Box h (B_0/m^2_p)^2$. Even at the Schwinger limit $B_c = m_e^2/q$, the correction is of order $\sim 10^{-90}$, far below anything detectable.

A complete treatment of photon-graviton mixing beyond tree-level can also be found in \cite{Bastianelli1, Bastianelli2}.

\section{Calculation of \texorpdfstring{$\varepsilon$}{Lg} and \texorpdfstring{$\mu$}{Lg} in the strong field regime}\label{appendix}
In order to compute the electric permittivity and magnetic permeability in (\ref{emu}), one has to compute the derivatives of (\ref{oneloop}) with respect to $a$ and $b$. This appendix presents a detailed calculation of these derivatives in the limit $a \ll 1$ and $b \gg 1$.

Let's start from the derivative with respect to $a$. In the limit $a \ll 1$, this is
\begin{equation}\label{A1}
    \frac{\partial \mathcal{L}_{EM}^{(1)}(a,b)}{\partial a} = \frac{a m_e^4}{12 \pi^2} \int_0^\infty \frac{e^{-s}}{s} \left( bs \coth{bs} - 1\right) ds.
\end{equation}
Defining $x\equiv bs$, one can write (\ref{A1}) as
\begin{equation}\label{A2}
    \frac{a m_e^4}{12 \pi^2} \int_0^\infty e^{-x/b} \left( \coth{x} - \frac{1}{x}\right) dx.
\end{equation}
Each of the two integrals in (\ref{A2}) diverges when evaluated separately, but if one regularizes the integrals by introducing a convergence-enforcing parameter $\epsilon$, the divergent parts are found to cancel out, leaving a finite result after summing.

In order to compute (\ref{A2}) we need the following standard integrals:
\begin{align}\label{tab1}
    \begin{split}
        &\int_0^\infty dx x^{m - 1} e^{-n x} = n^{-m} \Gamma(m), \\
        &\int_0^\infty dx x^m e^{-n x} \coth{x} = 2^{-m-1} \Gamma(1+m) \left[ \zeta\left(1+m, \frac{n}{2}\right) + \zeta\left(1+m, \frac{2+n}{2}\right)\right],
    \end{split}
\end{align}
where $\Gamma(x)$ is the gamma function, and $\zeta(x,y)$ the generalized Riemann zeta function. 

We regularize the integral (\ref{A2}) by multiplying both terms by $x^\epsilon$:
\begin{equation}\label{A4}
    \frac{a m_e^4}{12 \pi^2} \int_0^\infty \left( x^\epsilon e^{-x/b} \coth{x} - x^{\epsilon - 1} e^{-x/b} \right) dx.
\end{equation}
Using (\ref{tab1}), the integral in (\ref{A4}) evaluates to
\begin{equation}\label{A5}
\begin{split}
    &\int_0^\infty \left( x^\epsilon e^{-x/b} \coth{x} - x^{\epsilon - 1} e^{-x/b} \right) dx = \\
    &2^{-\epsilon-1} \Gamma(1+\epsilon) \left[ \zeta\left(1+\epsilon, \frac{1}{2b}\right) + \zeta\left(1+\epsilon, 1+\frac{1}{2b}\right)\right] - b^\epsilon \Gamma(\epsilon).
\end{split}
\end{equation}
In order to take the limit $\epsilon \rightarrow 0$, we need the following $\epsilon$ expansions:
\begin{align}
\begin{split}
    b^{\epsilon} &= 1 + \epsilon \ln b + \frac{1}{2} (\ln b)^2 \epsilon^2 + \mathcal{O}(\epsilon^3), \\
    \Gamma(\epsilon) &= \frac{1}{\epsilon} - \gamma + \frac{1}{12} (6 \gamma^2 + \pi^2 ) \epsilon + \mathcal{O}(\epsilon^2),\\
    \Gamma(1+\epsilon) &= 1 - \gamma \epsilon + \frac{1}{12} (6 \gamma^2 + \pi^2 ) \epsilon^2 + \mathcal{O}(\epsilon^3),\\
    \Gamma(\epsilon - 1) &= -\frac{1}{\epsilon} + \gamma -1 +\left( -\frac{\pi^2}{12} -\frac{\gamma^2}{2} - 1 + \gamma\right) \epsilon + \mathcal{O}(\epsilon^2), \\
    \zeta(\epsilon, h) &= \frac{1}{2} - h + \epsilon \zeta'(0,h) + \frac{\epsilon^2}{2} \zeta''(0,h) + \mathcal{O}(\epsilon^3), \\
    \zeta(1+\epsilon, h) &= \frac{1}{\epsilon} - \psi(h) + \mathcal{O}(\epsilon),\\
    \zeta(\epsilon - 1, h) &= -\frac{1}{2} h^2 +\frac{1}{2} h-\frac{1}{12} +\epsilon \zeta'(-1,h) +\frac{\epsilon^2}{2} \zeta''(-1,h) + \mathcal{O}(\epsilon^3),
\end{split}
\end{align}
where $\gamma$ is the Euler-Mascheroni constant and $\psi(z)$ the digamma function, defined as $\psi(x) \equiv \Gamma'(x)/\Gamma(x)$.

With the identities above, and in the limit $\epsilon \rightarrow 0$, (\ref{A5}) gives
\begin{equation}
    \lim_{\epsilon\rightarrow0} \int_0^\infty \left( x^\epsilon e^{-x/b} \coth{x} - x^{\epsilon - 1} e^{-x/b} \right) dx = - \psi\left(\frac{1}{2b}\right) - b - \ln(2b),
\end{equation}
where I have used the identity $\psi(1+x) = \psi(x) + 1/x$. As promised, the divergent pieces $\propto \epsilon^{-1}$ have canceled out.

Similarly, one can compute the derivative with respect to $b$ in the limit $a \ll 1$. This is
\begin{equation}
    \frac{\partial \mathcal{L}_{EM}^{(1)}(a,b)}{\partial b} = \frac{b m_e^4}{12 \pi^2} \int_0^\infty \frac{e^{-s}}{s} \left( 1 - \frac{3}{2}\frac{\coth(bs)}{bs} + \frac{3}{2} \csch^2(bs)\right) ds.
\end{equation}
With $x \equiv bs$, and introducing the regulator $x^\epsilon$, the integral becomes
\begin{equation}
    \frac{b m_e^4}{12 \pi^2} \lim_{\epsilon \rightarrow 0}\int_0^\infty e^{-x/b} \left( x^{\epsilon-1} - \frac{3}{2} x^{\epsilon - 2}\coth x + \frac{3}{2} x^{\epsilon-1}\csch^2 x \right) dx.
\end{equation}
Using (\ref{tab1}), we get
\begin{align}
\begin{split}
    &\frac{b m_e^4}{12 \pi^2} \Biggl[ - \gamma + \ln b -(1 - \gamma)\left( \frac{3}{4 b^2} + \frac{1}{2} \right) + \left( \frac{3}{4 b^2} + \frac{1}{2} \right) \ln2 + 6 \zeta'\left(-1,\frac{1}{2b}\right) - \frac{3}{2b} \ln(2b)  \\
    &- \gamma \left( \frac{3}{4 b^2} - \frac{1}{2} \right) - \left( \frac{3}{4 b^2} - \frac{1}{2} \right) \ln2 + 6 \zeta'\left(-1, \frac{1}{2b}\right) - \frac{3}{b} \ln \Gamma\left(\frac{1}{2b}\right) + \frac{3}{2b} \ln 2\pi\Biggl].
\end{split}
\end{align}

Therefore, the two derivatives are
\begin{align}\label{dLab}
\begin{split}
    \frac{\partial \mathcal{L}_{EM}^{(1)}(a,b)}{\partial a} &= \frac{a m_e^4}{12 \pi^2} \left( - \psi\left(\frac{1}{2b}\right) - b - \ln2b \right), \\
    \frac{\partial \mathcal{L}_{EM}^{(1)}(a,b)}{\partial b} &= \frac{b m_e^4}{12 \pi^2} \left( \ln 2b - \frac{3}{4b^2} + 12 \zeta'\left(-1,\frac{1}{2b}\right) + \frac{3}{2b} \ln \frac{\pi}{b} - \frac{3}{b} \ln \Gamma\left(\frac{1}{2b} \right) - \frac{1}{2} \right).
\end{split}
\end{align}
In the supercritical limit $b \gg 1$, $\psi(1/2b) \approx - 2b$ and $\Gamma(1/2b) \approx 2b$, therefore (\ref{dLab}) becomes
\begin{align}
\begin{split}
    \frac{\partial \mathcal{L}_{EM}^{(1)}(a,b)}{\partial a} &= \frac{a m_e^4}{12 \pi^2} \left( b - \ln b \right), \\
    \frac{\partial \mathcal{L}_{EM}^{(1)}(a,b)}{\partial b} &= \frac{b m_e^4}{12 \pi^2} \ln b.
\end{split}
\end{align}


\newpage
\begin{thebibliography}{}
\bibitem{Bohr}
Bohr, N., and Rosenfeld, L. (1933). On the question of the measurability of electromagnetic field quantities. Quantum theory and measurement, 479.
\bibitem{Dyson}
Dyson, F. (2013). Is a graviton detectable?. International Journal of Modern Physics A, 28(25), 1330041.
\bibitem{Bronstein}
Bronstein, M. P. (2005). Quantentheorie schwacher Gravitationsfelder, Phys. Ztschr. der Sowjetuion, 9, 140–157 (1936)
\bibitem{Feynman}
Feynman, R. (2018). Feynman lectures on gravitation. CRC Press.
\bibitem{Hannah}
Eppley, K., and Hannah, E. (1977). The necessity of quantizing the gravitational field. Foundations of Physics, 7, 51-68.
\bibitem{Page}
Page, D. N., and Geilker, C. D. (1981). Indirect evidence for quantum gravity. Physical Review Letters, 47(14), 979.
\bibitem{DePalma}
Mari, A., De Palma, G., and Giovannetti, V. (2016). Experiments testing macroscopic quantum superpositions must be slow. Scientific reports, 6(1), 22777.
\bibitem{Mattingly}
Mattingly, J. (2006). Why Eppley and Hannah’s thought experiment fails. Physical Review D, 73(6), 064025.
\bibitem{Baym}
Baym, G., and Ozawa, T. (2009). Two-slit diffraction with highly charged particles: Niels Bohr's consistency argument that the electromagnetic field must be quantized. Proceedings of the National Academy of Sciences, 106(9), 3035-3040.
\bibitem{Rydving}
Rydving, E., Aurell, E., and Pikovski, I. (2021). Do Gedanken experiments compel quantization of gravity?. Physical Review D, 104(8), 086024.
\bibitem{Belenchia}
Belenchia, A., Wald, R. M., Giacomini, F., Castro-Ruiz, E., Brukner, Č., and Aspelmeyer, M. (2018). Quantum superposition of massive objects and the quantization of gravity. Physical Review D, 98(12), 126009.
\bibitem{Danielson}
Danielson, D. L., Satishchandran, G., and Wald, R. M. (2022). Gravitationally mediated entanglement: Newtonian field versus gravitons. Physical Review D, 105(8), 086001.
\bibitem{Hossenfelder}
Hossenfelder, S. (2013). Minimal length scale scenarios for quantum gravity. Living Reviews in Relativity, 16(1), 2.
\bibitem{Calmet}
Calmet, X., Graesser, M., and Hsu, S. D. (2004). Minimum length from quantum mechanics and classical general relativity. Physical review letters, 93(21), 211101.
\bibitem{Krauss}
Krauss, L. M., and Wilczek, F. (2014). Using cosmology to establish the quantization of gravity. Physical Review D, 89(4), 047501.
\bibitem{Rothman}
Rothman, T., and Boughn, S. (2006). Can gravitons be detected?. Foundations of Physics, 36, 1801-1825.
\bibitem{Boughn}
Boughn, S., and Rothman, T. (2006). Aspects of graviton detection: graviton emission and absorption by atomic hydrogen. Classical and quantum gravity, 23(20), 5839.
\bibitem{Gertsenshtein}
Gertsenshtein, M. E. (1962). Wave resonance of light and gravitional waves. Sov Phys JETP, 14, 84-85.
\bibitem{Ejlli}
Ejlli, D., and Thandlam, V. R. (2019). Graviton-photon mixing. Physical Review D, 99(4), 044022.
\bibitem{Carney}
Carney, D., Domcke, V., and Rodd, N. L. (2024). Graviton detection and the quantization of gravity. Physical Review D, 109(4), 044009.
\bibitem{Anninos}
Anninos, P., Rothman, T., and Palessandro, A. (2024). Graviton-photon oscillations in an expanding universe. Physics of the Dark Universe, 44, 101480.
\bibitem{Pal}
Palessandro, A., and Rothman, T. (2023). A simple derivation of the Gertsenshtein effect. Physics of the Dark Universe, 40, 101187.
\bibitem{Zeldovich}
Zel’dovich, Y. B. (1973). Electromagnetic and gravitational waves in a stationary magnetic field. Zh. Eksp. Teor. Fiz, 65, 1311.
\bibitem{Weisskopf}
Weisskopf, V. (1936). On the electrodynamics of the vacuum on the basis of the quantum theory of the electron. Kong. Dans. Vid. Selsk. Mat. Fys. Medd, 14, 24.
\bibitem{Anninos2}
Anninos, P. (1998). Plane-symmetric cosmology with relativistic hydrodynamics. Physical Review D, 58(6), 064010.
\bibitem{Tsagas}
Tsagas, C. G. (2004). Electromagnetic fields in curved spacetimes. Classical and Quantum Gravity, 22(2), 393.
\bibitem{Lifshitz}
Berestetskii, V. B., Lifshitz, E. M., and Pitaevskii, L. P. (1982). Quantum Electrodynamics: Volume 4 (Vol. 4). Butterworth-Heinemann.
\bibitem{Domcke}
Domcke, V., and Garcia-Cely, C. (2021). Potential of radio telescopes as high-frequency gravitational wave detectors. Physical review letters, 126(2), 021104.
\bibitem{EH}
Heisenberg, W., and Euler, H. (1936). Folgerungen aus der diracschen theorie des positrons. Zeitschrift für Physik, 98(11-12), 714-732.
\bibitem{Dittrich}
Dittrich, W., and Gies, H. (2000). Probing the quantum vacuum: perturbative effective action approach in quantum electrodynamics and its application (Vol. 166). Springer Science and Business Media.
\bibitem{Lundin}
Lundin, J. (2010). QED and collective effects in vacuum and plasmas (Doctoral dissertation, Umeå universitet. Institutionen för fysik).
\bibitem{Kim}
Kim, C. M., and Kim, S. P. (2021). Magnetars as laboratories for strong field QED. arXiv preprint arXiv:2112.02460.
\bibitem{Medvedev}
Medvedev, M. V. (2023). Plasma modes in QED super-strong magnetic fields of magnetars and laser plasmas. Physics of Plasmas, 30(9).
\bibitem{CAST}
Anastassopoulos, V., A. Liolios, J. I. Collar, C. J. Hailey, H. Bräuninger, Igor Garcia Irastorza, A. Dermenev et al. "arXiv: New CAST Limit on the Axion-Photon Interaction." Nature Phys. 13, no. arXiv: 1705.02290 (2017): 584-590.
\bibitem{Natarajan}
Natarajan, C. M., Tanner, M. G., and Hadfield, R. H. (2012). Superconducting nanowire single-photon detectors: physics and applications. Superconductor science and technology, 25(6), 063001.
\bibitem{Bai}
Bai, P., Zhang, Y. H., and Shen, W. Z. (2017). Infrared single photon detector based on optical up-converter at 1550 nm. Scientific Reports, 7(1), 15341.
\bibitem{Pal2}
Palessandro, A., and Sloth, M. S. (2020). Gravitational absorption lines. Physical Review D, 101(4), 043504.
\bibitem{Martin}
Martin, J., and Brandenberger, R. H. (2001). Trans-Planckian problem of inflationary cosmology. Physical Review D, 63(12), 123501.
\bibitem{Niemeyer}
Niemeyer, J. C. (2001). Inflation with a Planck-scale frequency cutoff. Physical Review D, 63(12), 123502.
\bibitem{TPCC}
Bedroya, A., and Vafa, C. (2020). Trans-Planckian censorship and the swampland. Journal of High Energy Physics, 2020(9), 1-34.
\bibitem{Bedroya}
Bedroya, A., Brandenberger, R., Loverde, M., and Vafa, C. (2020). Trans-Planckian censorship and inflationary cosmology. Physical Review D, 101(10), 103502.
\bibitem{CCC}
Penrose, R. (1969). Gravitational collapse: The role of general relativity. Nuovo Cimento Rivista Serie, 1, 252.
\bibitem{Penrose}
Penrose, R. (1973). Naked singularities. Annals of the New York Academy of Sciences, 224(1), 125-134.
\bibitem{Brahma}
Brahma, S. (2020). Trans-Planckian censorship, inflation, and excited initial states for perturbations. Physical Review D, 101(2), 023526.
\bibitem{Jacobson}
Jacobson, T. (1995). Thermodynamics of spacetime: the Einstein equation of state. Physical Review Letters, 75(7), 1260.
\bibitem{Padmanabhan}
Padmanabhan, T. (2010). Thermodynamical aspects of gravity: new insights. Reports on Progress in Physics, 73(4), 046901.
\bibitem{Verlinde}
Verlinde, E. (2011). On the origin of gravity and the laws of Newton. Journal of High Energy Physics, 2011(4), 1-27.
\bibitem{AdS/CFT}
Maldacena, J. (1999). The large-N limit of superconformal field theories and supergravity. International journal of theoretical physics, 38(4), 1113-1133.
\bibitem{Jentschura}
Jentschura, U. D., Gies, H., Valluri, S. R., Lamm, D. R., and Weniger, E. J. (2002). QED effective action revisited. Canadian Journal of Physics, 80(3), 267-284.
\bibitem{Dunne}
Dunne, G. V. (2005). Heisenberg–Euler effective Lagrangians: basics and extensions. In From Fields to Strings: Circumnavigating Theoretical Physics: Ian Kogan Memorial Collection (In 3 Volumes) (pp. 445-522).
\bibitem{Chadha}
Chadha, S., and Olesen, P. (1977). On Borel singularities in quantum field theory. Physics Letters B, 72(1), 87-90.
\bibitem{Schwinger}
Schwinger, J. (1951). On gauge invariance and vacuum polarization. Physical Review, 82(5), 664.
\bibitem{Ruhdorfer}
Ruhdorfer, M., Serra, J., and Weiler, A. (2020). Effective field theory of gravity to all orders. Journal of High Energy Physics, 2020(5), 1-36.
\bibitem{Bastianelli1}
Bastianelli, F., and Schubert, C. (2005). One loop photon-graviton mixing in an electromagnetic field: Part 1. Journal of High Energy Physics, 2005(02), 069.
\bibitem{Bastianelli2}
Bastianelli, F., Nucamendi, U., Schubert, C., and Villanueva, V. M. (2007). One loop photon-graviton mixing in an electromagnetic field: part 2. Journal of High Energy Physics, 2007(11), 099.

\end{thebibliography}
\end{document}